\documentclass[conference]{IEEEtran}
\IEEEoverridecommandlockouts
\usepackage{cite}
\usepackage{amsmath,amssymb,amsfonts}
\usepackage[utf8]{inputenc}
\usepackage{xcolor}
\usepackage{listings}
\usepackage{graphicx}
\usepackage{tikz}
\usepackage{pgf}
\usepackage{float}
\usepackage{booktabs}
\usepackage{pgfplots}
\usepackage{algorithm}
\usepackage{algpseudocode}
\pgfplotsset{compat=1.18}
\usepgfplotslibrary{groupplots}
\usepackage{textcomp}
\usepackage[capitalise]{cleveref}

\usepackage{algpseudocode}
\usepackage{algorithm}
\usepackage{xcolor}
\usepackage{tcolorbox}
\usepackage{booktabs}
\usepackage[inline,shortlabels]{enumitem}

\usepackage{pgfplots}
\usepackage{pgfplotstable}
\usepackage{tikz}
\usepackage{subcaption}
\usepackage{caption}
\usepgfplotslibrary{dateplot,statistics,groupplots}
\usetikzlibrary{patterns} 
\pgfplotsset{compat=1.18}

\definecolor{keywordblue}{rgb}{0.0, 0.33, 0.55} 
\definecolor{commentgreen}{rgb}{0.0, 0.6, 0.0}  
\definecolor{stringred}{rgb}{0.75, 0.0, 0.0}    
\definecolor{decoratorpurple}{rgb}{0.5, 0.0, 0.5} 

\lstdefinelanguage{CompoundAI}{
    morekeywords={
        Structure, Function, List, String, Float, Integer, Image,
        task_type, parameters, task_slos, return, pass, str, float,
        int, list, tuple, def, class
    },
    classoffset=1, 
    morekeywords={@CompoundableTask, @DataContract, @TaskContract, TaskType, SLO, ObjectDetection, DataContract},
    keywordstyle=\bfseries\color{decoratorpurple},
    classoffset=0,
    sensitive=true,
    morecomment=[l]{\#}, 
    morestring=[b]", 
}

\usepackage{xcolor}
\def\BibTeX{{\rm B\kern-.05em{\sc i\kern-.025em b}\kern-.08em
    T\kern-.1667em\lower.7ex\hbox{E}\kern-.125emX}}
\begin{document}
\bstctlcite{BSTcontrol}
\pagestyle{empty}

\title{\huge PLAIground: SLO-Driven Runtime Model Selection for Compound AI Systems in the Edge-Cloud-Space Continuum}

\author{
\IEEEauthorblockN{Milos Gravara}
\IEEEauthorblockA{
Distributed Systems Group \\
TU Wien \\
m.gravara@dsg.tuwien.ac.at
}
\and
\IEEEauthorblockN{Cynthia Marcelino}
\IEEEauthorblockA{
Distributed Systems Group \\
TU Wien \\
c.marcelino@dsg.tuwien.ac.at
}
\and
\IEEEauthorblockN{Andrija Stanisic}
\IEEEauthorblockA{
Distributed Systems Group \\
TU Wien \\
a.stanisic@dsg.tuwien.ac.at
}
\and
\IEEEauthorblockN{Stefan Nastic}
\IEEEauthorblockA{
Distributed Systems Group \\
TU Wien \\
s.nastic@dsg.tuwien.ac.at
}
}

\maketitle

\begin{abstract}

Applications in the 3D Computing Continuum, which unifies edge, cloud, and space, require combining multiple AI tasks such as object detection, time-series analytics, and natural language processing into Compound AI systems. These systems must satisfy stringent Service Level Objectives (SLOs) on accuracy, latency, and cost. A key mechanism for maintaining SLO compliance of Compound AI systems is runtime model selection, where AI models are dynamically switched for each workflow task. However, existing distributed and compound AI frameworks do not natively support runtime model selection. We present PLAIground, a framework that enables runtime model selection for Compound AI systems. PLAIground introduces Compoundable AI Model (CAIM) abstraction, which decouples task semantics from AI model implementations via Task and Data Contracts, enabling model switching without workflow changes. Additionally, PLAIground introduces Pixie, an SLO-driven runtime model selection algorithm, which dynamically selects the most suitable model for each task during execution. Our evaluation on two realistic Compound AI workflows demonstrates that Pixie achieves up to 91.3\% accuracy while maintaining SLO compliance where fixed-model strategies either violate cost and latency budgets up to 21x or miss accuracy targets by 4\%.
\end{abstract}


\section{Introduction}
\label{sec:introduction}

The emerging 3D Computing Continuum~\cite{HyperDrive2024} unifies Edge, Cloud, and Space, transforming LEO satellites into onboard processing nodes that move computation closer to the data source. Applications in this environment require intelligent processing no single model can provide~\cite{Gravara2025ANC}. For example, wildfire detection combines multiple AI tasks, such as object detection on satellite imagery, time-series analytics on sensor streams and natural language processing for emergency alerts. Such applications naturally decompose into Compound AI systems - distributed AI systems that orchestrate multiple AI models, tools and specialized algorithms into workflows~\cite{Gravara2025ANC, chen2025optimizingmodelselectioncompound}. Compound AI systems operating in the 3D Computing Continuum must satisfy strict Service Level Objectives (SLOs) on accuracy, latency, and cost (e.g. energy consumption). Maintaining SLO compliance is particularly challenging in this environment, as conditions change continuously: satellite coverage shifts, available bandwidth fluctuates, and energy budgets deplete~\cite{HyperDrive2024, Databelt}.

One of the key mechanisms for achieving SLO compliance in Compound AI systems is model selection~\cite{chen2025optimizingmodelselectioncompound, EdgeAdaptor, COMPASS}. Each task in a Compound AI workflow can be fulfilled by multiple candidate AI models. Selecting the most suitable AI model for each task allows the system to maintain SLO compliance as conditions change~\cite{EdgeAdaptor, COMPASS}. This is especially relevant in the 3D Computing Continuum, where resource constraints limit horizontal scaling. Instead of provisioning additional infrastructure, adaptation can be achieved by switching between different AI models for each task at runtime, as conditions and observed performance change. Prior work on inference serving has shown that dynamically selecting model variants can reduce costs by up to 10 $\times$ over static deployments~\cite{INFaaS}, though only for single-model inference serving.

However, Compound AI systems make this problem considerably more challenging. Multiple candidate AI models exist for each task, and each exhibits different trade-offs between accuracy, latency, and cost~\cite{chen2025optimizingmodelselectioncompound, OPTSURVEY, INFaaS}. For instance, a lightweight object detector runs efficiently on resource-constrained devices but yields lower accuracy, while a larger model provides higher accuracy at greater computational cost. With $n$ tasks and $m$ candidates per task, a Compound AI workflow theoretically has $m^n$ possible AI model selections. In order to maintain SLO compliance, this space must be navigated continuously, as the most suitable configuration changes when conditions change.

This challenge is further complicated by the limitations of existing Compound AI frameworks, which do not provide built-in support for runtime model selection. Frameworks such as LangChain~\cite{chase2022langchain} and DSPy~\cite{DSPY} couple task definitions to specific AI model implementations, i.e., each AI task explicitly specifies which model to invoke and how to parse its responses. However, various models have different input/output formats, API interfaces and deployment configurations. Switching from one model to another requires modifying application code, rewriting API calls, and adjusting data handling. There is no separation between what a task performs and which AI model performs it. 

In this paper, we present PLAIground, a framework that enables SLO-driven runtime model selection for Compound AI systems through novel abstractions that decouple workflow logic from specific AI models and runtime mechanisms that maintain SLO compliance through continuous model assignment.

Our main contributions include:

\begin{enumerate}
    \item[\textbf{C-1:}] \textbf{Compoundable AI Model (CAIM):} A novel programming abstraction that encapsulates Compound AI task semantics through Task and Data Contracts, decoupling workflow logic from specific AI models and providing a foundation for runtime model selection. 
    \item[\textbf{C-2:}] \textbf{Pixie: SLO-Driven Runtime Model Selection Algorithm:} A novel runtime model selection algorithm that continuously assigns the most suitable AI model to each CAIM, while ensuring SLO compliance. Evaluation shows that Pixie is the only approach that maintains SLO compliance across two Compound AI workflows, while achieving up to 91.3\% accuracy, and reducing costs by up to 21x and latency by 2.5x, compared to static baselines optimizing for a single metric.
\end{enumerate}
\section{Illustrative Scenario and Research Questions}\label{sec:motivation}

To motivate the challenges of enabling SLO-driven runtime model selection, we present an illustrative scenario and our main research questions.

\subsection{Main Illustrative Scenario}

\begin{figure}[h]
    \centering
    \includegraphics[width=0.6\linewidth]{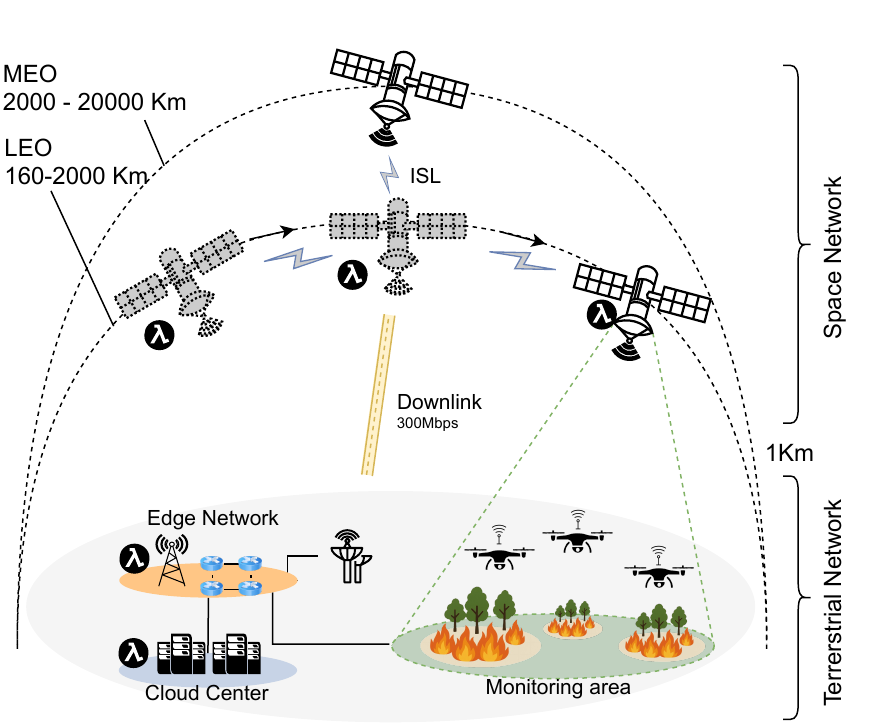}
    \caption{Wildfire Illustrative Scenario in the 3D Continuum}
    \label{fig:scenario}
\end{figure}

Early detection of wildfires in remote areas requires low-latency data processing and efficient use of heterogeneous hardware resources across the 3D Computing Continuum. In our wildfire scenario (\cref{fig:scenario}), drones capture video and sensor data, streaming to nearby edge nodes or LEO satellites when ground coverage is unavailable. This data flows through a Compound AI workflow consisting of object detection, data analytics, and alarm generation. Each task can be fulfilled by multiple candidate AI models with different performance characteristics. For object detection alone, options range from lightweight models like YOLOv5n for resource-constrained environments to high-accuracy models like YOLOv8x requiring substantial compute.

These workflows operate under SLOs that define acceptable bounds for accuracy, latency, and cost. Consider the object detection task running on a LEO satellite with a limited energy budget. Initially, YOLOv8x satisfies both accuracy and energy SLOs. As the satellite's battery depletes, YOLOv8x becomes too costly to run and violates the energy constraint. A switch to the lighter YOLOv5n preserves the energy SLO, at the cost of reduced detection accuracy. The most suitable model depends on current conditions, and those conditions change continuously throughout the workflow execution.



However, runtime model switching is far from straightforward. The workflow integrates YOLOv8x through model-specific code for API invocation, input preprocessing, and output parsing. Switching to YOLOv5n requires the developer to have anticipated this scenario, written integration code for both models, and built the switching logic into the application. With multiple tasks in the workflow and multiple candidate models per task, each with different interfaces and deployment requirements, this manual approach does not scale to dynamic environments such as the 3D Computing Continuum.


\subsection{Research Questions}

We identify following main research questions:

\noindent\emph{\textbf{RQ-1:} How to design abstractions that decouple Compound AI workflow specification from underlying AI models and, thus, enable runtime model selection?}


Even when performing same task, different AI models enforce incompatible data interfaces, ranging from raw tensors in hardware optimized runtimes (e.g. Triton) to JSON payloads for proprietary APIs. This causes tight coupling that forces developers to hardcode model-specific logic into the workflow. Enabling runtime switching requires novel abstractions which decouple task semantics from implementation details. 


    


\noindent\textit{\textbf{RQ-2}: How to maintain SLO compliance of Compound AI systems through selection of most suitable models for each task at runtime?} 


Given a set of candidate models per task, each candidate exhibits different trade-offs between accuracy, latency, and cost, and the most suitable AI model shifts as conditions change. Selecting too conservatively sacrifices accuracy, while selecting too aggressively risks SLO violations. Navigating these trade-offs across all tasks in a workflow, under continuously changing conditions, necessitates dedicated runtime mechanisms that monitor performance and adapt model assignments accordingly.
    



\section{Compoundable AI Model}\label{sec:model}



\begin{figure}[h]
  \centering
  \includegraphics[width=0.7\linewidth]{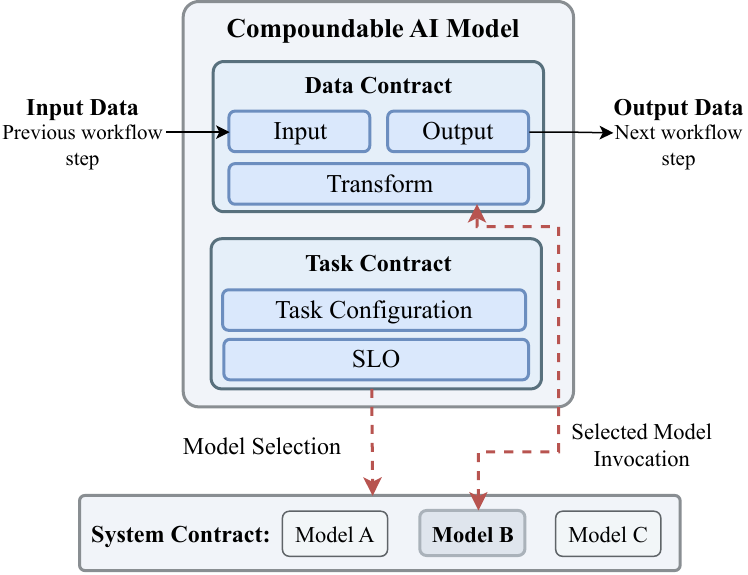}
  \caption{Compoundable AI Model Abstraction}
  \label{fig:programming-model}
\end{figure}

The Compoundable AI Model (CAIM), presented in \cref{fig:programming-model}, is a programming abstraction representing the main building block of Compound AI Workflows. Its main purpose is to facilitate the development of Compound AI workflows and enable the runtime model selection mechanism without modifying the workflow itself. Consequently, this allows developers to focus on the design of Compound AI applications, delegating execution details of model selection entirely to the runtime. For example, a developer can specify that a certain workflow step needs to perform object detection for wildfires under SLO constraints; the underlying system then finds a set of candidate AI models that can perform the given step and selects the most suitable AI model for execution. When the performance drifts or resources fluctuate, the runtime can switch between the candidate models to maintain the SLO compliance.

The CAIM abstraction defines three contracts: 1) \textit{Task Contract}, which mainly specifies functional and non-functional requirements; 2) \textit{Data Contract}, defining input/output interfaces; and 3) \textit{System Contract}, which contains the candidate AI models capable of executing the specified task, along with their performance profiles and deployment specifications. In the following, we detail the Task and Data Contract as the developer-specified components of the CAIM. System Contract is platform-provided and contains the candidate AI models eligible for runtime selection, along with their performance profiles and deployment specifications. Producing the System Contract from a broader model registry is orthogonal to this work~\cite{chen2025optimizingmodelselectioncompound, OPTSURVEY}.

\subsection{Task Contract}

The Task Contract is a declarative specification that defines what AI task a CAIM is expected to perform, independent of which AI model implementation performs it. It combines functional requirements, specifying the task to be executed, with non-functional requirements, specifying the constraints under which execution must occur. 

Functional requirements define what type of AI task CAIM is expected to perform and its task-specific configurations. The task type identifies the required capability (e.g., object detection, text generation), while configuration parameters capture task-specific settings such as target classes for detection, or prompt templates for language models. Together, these establish the capability requirements that any candidate AI model must satisfy. For example, specifying object detection with classes \texttt{[fire, smoke]} means only models capable of detecting these classes qualify as candidates.

Non-functional requirements specify SLOs that constrain how the task must be performed. The Task Contract distinguishes two categories: Task SLOs define minimum quality requirements on the output, establishing a quality floor that all candidate AI models must satisfy (e.g., minimum accuracy, precision, or recall thresholds). System SLOs define efficiency constraints on execution, such as maximum latency, monetary cost, or energy consumption. System SLOs can be specified directly per CAIM, or at the workflow level (e.g., total cost budget, end-to-end latency), in which case they are decomposed into per-CAIM budgets at deployment time (detailed in \cref{subsec:pixie}).




\subsection{Data Contract}

The Data Contract defines the strict input and output schemas for a CAIM. Each contract declares specific data types, establishing clear interfaces that allow CAIMs to connect into workflows regardless of the underlying AI model executing the task. The schema specification supports simple types as well as complex structures like nested objects, arrays, and domain-specific types such as bounding boxes.

Consequently, the Data Contract acts as a normalization layer. Since different underlying AI models may produce outputs in varying formats (e.g., raw tensors vs. JSON strings), the abstraction ensures that all execution outputs are adapted to match the defined Data Contract. This guarantees that runtime model selection does not break the workflow. This means that when AI models are switched dynamically, downstream tasks always receive data in the format defined by the Data Contract.



Compound AI workflows are constructed by defining CAIMs as computational steps connected through this explicit data flow. Each step is declared by its Task Contract, with Data Contracts creating the stable interfaces for the complete workflow logic. The workflow logic remains independent of which AI models execute each step, as Task Contracts decouple task semantics from model implementations and Data Contracts guarantee interface
consistency. This enables runtime model selection without modifying workflow logic, thereby addressing \textbf{RQ-1}.
\section{Pixie: SLO-Driven Runtime Model Selection}\label{subsec:pixie}

Pixie is a runtime model selection algorithm that maximizes accuracy while maintaining SLO compliance. It operates independently per CAIM, selecting and adapting AI model assignments during workflow execution, based on the System Contract and the System SLOs. Pixie's design is grounded in the monotonicity property observed in Compound AI systems~\cite{chen2025optimizingmodelselectioncompound}: improving accuracy at any individual workflow step does not degrade end-to-end workflow accuracy. Under this property, independent per-CAIM accuracy maximization is sufficient for maximizing workflow-level accuracy. \cref{alg:pixie} presents Pixie. 

Initially, Pixie greedily selects the highest-accuracy candidate whose profiled metrics satisfy System SLOs. During execution, Pixie monitors observed resource metrics (e.g., latency, cost, energy) through windowed moving averages. After each request, the observed metrics update the windows (line~\ref{alg:pixie:update}). For each System SLO, Pixie computes the normalized gap between the SLO limit and the observed average. The minimum gap across all System SLOs identifies the most constrained resource (line~\ref{alg:pixie:gap}). Two configurable thresholds, $\tau_{low}$ and $\tau_{high}$, define three operating zones. If the minimum gap for most constrained resource falls below $\tau_{low}$, its System SLO is under pressure, and Pixie downgrades to a less resource-intensive candidate (line~\ref{alg:pixie:downgrade}). If the minimum gap exceeds $\tau_{high}$, all System SLOs have sufficient headroom, and Pixie upgrades to a higher-accuracy candidate (line~\ref{alg:pixie:upgrade}). Between these thresholds, the current assignment remains unchanged. If no further downgrade or upgrade is available, execution continues with the current AI model. After each model selection decision, Pixie resets the observation windows and enters a cooldown period of $k$ observations before re-evaluating (line~\ref{alg:pixie:cooldown}). This prevents oscillation from noisy measurements after a transition.

\begin{algorithm}[h]
\caption{Pixie: SLO-Aware Runtime Model Selection}
\label{alg:pixie}
\scriptsize
\begin{algorithmic}[1]
\Require Ordered candidate set $\mathcal{C}$ with profiles \Comment{System Contract}
\Require System SLOs $\mathcal{S} = \{(R_i, L_i)\}_{i=1}^{n}$ \Comment{Task Contract}
\Require Window size $k$, thresholds $\tau_{low}$, $\tau_{high}$

\State $model \gets \Call{SelectInitial}{\mathcal{C}, \mathcal{S}}$ \Comment{Highest accuracy within SLO bounds}
\label{alg:pixie:init}
\State $W \gets \Call{InitWindows}{k}$

\While{workflow is running}
    \State $d \gets \Call{NextRequest}{}$

    \If{$\Call{WindowReady}{W}$} \Comment{Cooldown: skip until $k$ observations}
        \label{alg:pixie:cooldown}
        \State $g \gets \min_{i}\; (L_i - \Call{Avg}{W, R_i}) \;/\; L_i$ \Comment{Most constrained SLO}
        \label{alg:pixie:gap}

        \If{$g < \tau_{low}$} \Comment{SLO pressure}
            \label{alg:pixie:violation}
            \State $model \gets \Call{Downgrade}{model, \mathcal{C}}$
            \label{alg:pixie:downgrade}
            \State $\Call{Reset}{W}$

        \ElsIf{$g > \tau_{high}$} \Comment{SLO headroom}
            \label{alg:pixie:headroom}
            \State $model \gets \Call{Upgrade}{model, \mathcal{C}}$
            \label{alg:pixie:upgrade}
            \State $\Call{Reset}{W}$
        \EndIf
    \EndIf

    \State $result, \mathbf{o} \gets \Call{Execute}{d, model}$ \Comment{$\mathbf{o}$: observed metrics}
    \label{alg:pixie:exec}
    \State $\Call{Update}{W, \mathbf{o}}$
    \label{alg:pixie:update}
\EndWhile

\end{algorithmic}
\end{algorithm}

As System SLOs on cumulative resources (e.g., total cost, end-to-end latency) are typically specified at the workflow level, PLAIground decomposes them into per-CAIM budgets at deployment time. Each CAIM receives a budget share proportional to the average resource consumption of its candidates relative to the workflow total, ensuring that CAIMs whose candidates are intrinsically more resource-intensive receive proportionally larger budget shares. Decomposition of Task SLOs (e.g., mapping workflow-level accuracy to per-CAIM accuracy) is essentially an accuracy estimation problem, which remains an open challenge in Compound AI systems~\cite{OPTSURVEY}. In this paper, we focus on accuracy maximization under SLO constraints and leave Task SLO decomposition as future work.


Through continuous runtime model selection, Pixie maintains SLO compliance while maximizing task accuracy, thus answering \textbf{RQ-2}.

\section{Evaluation}\label{sec:evaluation}

This section evaluates PLAIground through two Compound AI workflows that represent different application domains and operational constraints: wildfire detection and question answering. PLAIground is implemented in Python as an open-source prototype\footnote{\label{fn:repo}https://github.com/polaris-slo-cloud/PLAIground}. We evaluate Pixie's runtime model selection on each workflow individually. 




\subsection{Experimental Setup}

\subsubsection{Workflows}

\textbf{Wildfire Detection.} A distributed pipeline for early wildfire detection in remote areas, composed of two CAIMs: (1) object detection on edge hardware (e.g., satellite) to identify fire and smoke in captured imagery, and (2) ground-based alert generation using LLMs to produce emergency notifications. Our evaluation focuses on the object detection stage, which operates under energy constraints. 


\textbf{QARouter.} A conditional routing workflow for question answering consisting of three CAIMs: a difficulty classifier that routes queries based on complexity, and two QA solvers (Simple QA and Complex QA) that handle easy and hard questions respectively. 


\subsubsection{Baselines}

Table~\ref{tab:baselines} presents the model selection strategies evaluated in our experiments.

\begin{table}[h]
\centering
\caption{Model selection strategies.}
\label{tab:baselines}
\resizebox{\linewidth}{!}{%
\begin{tabular}{lll}
\toprule
\textbf{Strategy} & \textbf{Description} & \textbf{Selection Mechanism} \\
\midrule
Random & Randomly from registry. & Baseline (no optimization) \\
Greedy-Cost & Selects lowest-cost & Static cost minimization \\
Greedy-Latency & Selects fastest (p95) & Static latency optimization \\
Greedy-Quality & Selects highest-accuracy & Static quality maximization \\
Pixie (Adaptive) & SLO-based dynamic switching. & Runtime adaptation \\
\bottomrule
\end{tabular}
}
\end{table}

For Wildfire Detection, Greedy-Cost optimizes energy consumption (Joules), while for QARouter optimizes for monetary cost (\$). Prior approaches~\cite{EdgeAdaptor, chen2025optimizingmodelselectioncompound} address deployment-time optimization for specific domains, while our focus is online adaptation during workflow execution. We, therefore, evaluate against algorithmic baselines representing the optimization targets (cost, latency, accuracy) that Pixie balances dynamically.

\begin{figure*}[t!]
    \begin{subfigure}{0.2\linewidth}
        \begin{tikzpicture}
            \begin{axis}[
                boxplot/draw direction=y,
                width=4.5cm, height=3.5cm,
                ymajorgrids=true, grid style=dashed,
                xtick={1,2,3,4,5,6},
                xlabel={~},
                ymin=0.6,
                ymax=1.0,
                xticklabels={Greedy(L),Greedy(C),Random,Greedy(Q), Pixie},
                xticklabel style={font=\tiny, rotate=25, anchor=north east},
                ylabel style={font=\footnotesize,yshift=-5pt},
                yticklabel style={font=\footnotesize},
                enlarge x limits=0.15,
                ylabel={Accuracy}
            ]
        
                \addplot+[
                    boxplot prepared={
                        median=0.7784,
                        lower quartile=0.7761,
                        upper quartile=0.7807,
                        lower whisker=0.7722,
                        upper whisker=0.7825,
                        draw position=1
                    },
                    fill=orange!40,
                    draw=orange!80,
                ] coordinates {(1,0)};
            
                \addplot+[
                    boxplot prepared={
                        median=0.76,
                        upper quartile=0.76,
                        lower quartile=0.76,
                        upper whisker=0.76,
                        lower whisker=0.76,
                        draw position=2
                    },
                    fill=green!60!black!40,
                    draw=green!60!black!80,
                ] coordinates {(2,0)};
                
                \addplot+[
                    boxplot prepared={
                        median=0.6958,
                        upper quartile=0.7339,
                        lower quartile=0.6577,
                        upper whisker=0.8215,
                        lower whisker=0.6326,
                        draw position=3
                    },
                    fill=purple!40,
                    draw=purple!80,
                ] coordinates {(3,0)};
            
                \addplot+[
                    boxplot prepared={
                        median=0.9344,
                        upper quartile=0.9412,
                        lower quartile=0.9311,
                        upper whisker=0.944,
                        lower whisker=0.9301,
                        draw position=4
                    },
                    fill=blue!40,
                    draw=blue!80,
                ] coordinates {(4,0)};

                \addplot+[
                    boxplot prepared={
                        median=0.8771,
                        upper quartile=0.8871,
                        lower quartile=0.8663,
                        upper whisker=0.891,
                        lower whisker=0.8600,
                        draw position=5
                    },
                    fill=red!40,
                    draw=red!80,
                ] coordinates {(5,0)};
        
            \end{axis}
        \end{tikzpicture}
        \caption{Accuracy Distribution}
        \label{fig:accuracy_boxplot}
    \end{subfigure}
    \begin{subfigure}{0.34\linewidth}
        \begin{tikzpicture}
        \begin{axis}[
          ybar,
          width=6.5cm, height=3.5cm,
          bar width=6pt,
          enlarge x limits=0.12,
          ymin=0.5, ymax=1,
          ylabel={Accuracy},
          ymajorgrids=true, grid style=dashed,
          ylabel style={font=\footnotesize,yshift=-5pt},
          yticklabel style={font=\footnotesize},
          legend style={at={(0.5,1.05)}, anchor=south, draw=none, legend columns=3, font=\footnotesize},
          tick align=inside,
          symbolic x coords={
              Greedy(L),sep1,sep1b,
              Greedy(C),sep2,sep2b,
              Random,sep3,sep3b,
              Greedy(Q),sep4,sep4b,
              Pixie,sep5,sep5b
            },
            xtick=data,
            xticklabels={,,,,,,}, 
            extra x ticks={Greedy(L),Greedy(C),Random,Greedy(Q),Pixie},
            extra x tick labels={Greedy(L),Greedy(C),Random,Greedy(Q),Pixie},
            extra x tick style={tick label style={font=\tiny, rotate=25,anchor=north east}}
        ]
        
        \pgfplotsset{
          leftshift/.style ={bar shift=-1.1*\pgfplotbarwidth},
          midshift/.style  ={bar shift= 0pt},
          rightshift/.style={bar shift=+1.1*\pgfplotbarwidth},
        }
        
        \addplot+[
        fill=blue!60, leftshift, error bars/.cd, y dir=both, y explicit
        ]
        coordinates {
          (Greedy(L),0.7784) +- (0,0.0035)
          (sep1,nan) (sep1b,nan)
          (Greedy(C),0.76)   +- (0,0)
          (sep2,nan) (sep2b,nan)
          (Random,0.6958)    +- (0,0.0738)
          (sep3,nan) (sep3b,nan)
          (Greedy(Q),0.9344) +- (0,0.0063)
          (sep4,nan) (sep4b,nan)
          (Pixie,0.8771)  +- (0,0.005)
          (sep5,nan) (sep5b,nan)
        };
        
        \addplot+[fill=green!60, draw=green!60!black, midshift, error bars/.cd, y dir=both, y explicit,
        error mark options={draw=green!60!black},  
        error mark=|, 
        ]
        coordinates {
          (Greedy(L),0.7839) +- (0,0.0062)
          (sep1,nan) (sep1b,nan)
          (Greedy(C),0.7231) +- (0,0)
          (sep2,nan) (sep2b,nan)
          (Random,0.7274)    +- (0,0.0648)
          (sep3,nan) (sep3b,nan)
          (Greedy(Q),0.9384) +- (0,0.0007)
          (sep4,nan) (sep4b,nan)
          (Pixie,0.856)   +- (0,0.002)
          (sep5,nan) (sep5b,nan)
        };
        
        \addplot+[fill=red!60, rightshift, error bars/.cd, y dir=both, y explicit]
        coordinates {
          (Greedy(L),0.7679) +- (0,0.0072)
          (sep1,nan) (sep1b,nan)
          (Greedy(C),0.8286) +- (0,0)
          (sep2,nan) (sep2b,nan)
          (Random,0.6465)    +- (0,0.0926)
          (sep3,nan) (sep3b,nan)
          (Greedy(Q),0.9261) +- (0,0.0044)
          (sep4,nan) (sep4b,nan)
          (Pixie, 0.8865) +- (0,0.0011)
          (sep5,nan) (sep5b,nan)
        };
        
        \legend{Overall, Easy, Hard}
        \end{axis}
        \end{tikzpicture}
        \vspace{0.7em}
    \caption{Accuracy Breakdown}
    \label{fig:accuracy_comparison}
\end{subfigure}
    \begin{subfigure}{0.22\linewidth}
        \begin{tikzpicture}
            \begin{axis}[
                ybar,
                symbolic x coords={Greedy(L), Greedy(C), Random, Greedy(Q), Pixie},
                xtick=data,
                ymin=500,
                ylabel={Latency (ms)},
                xlabel={~},
                xticklabel style={font=\tiny, rotate=25,anchor=north east},
                ylabel style={font=\footnotesize,yshift=-5pt},
                yticklabel style={font=\footnotesize}, 
                bar width=10pt,
                width=4.5cm,         
                height=3.5cm,
                enlarge x limits=0.2,
                ymajorgrids=true,
                tick align=inside,
                grid style=dashed,
            ]
                \addplot[fill=orange!50, error bars/.cd, y dir=both,
                    y explicit] coordinates {
                    (Greedy(L),651.08) +- (0,0)
                    (Greedy(C),671.64) +- (0,57.32)
                    (Random,1740.07) +- (0,419.20)
                    (Greedy(Q),2380.43) +- (0,98.72)
                    (Pixie, 906.99) +- (0, 95.34)
                };

            \end{axis}
        \end{tikzpicture}
        \caption{P95 Latency}
        \label{fig:latency_comparison}
    \end{subfigure}
    \begin{subfigure}{0.2\linewidth}
        \begin{tikzpicture}
            \begin{axis}[
                ybar,
                symbolic x coords={Greedy(L), Greedy(C), Random, Greedy(Q), Pixie},
                xtick=data,
                ymin=0,
                ylabel={Cost (USD)},
                xlabel={~},
                xticklabel style={font=\tiny, rotate=25, anchor=north east},
                ylabel style={font=\footnotesize,yshift=-5pt},
                yticklabel style={font=\footnotesize}, 
                bar width=10pt,
                width=4.5cm,         
                height=3.5cm,
                enlarge x limits=0.2,
                tick align=inside,
                ymajorgrids=true,
                grid style=dashed,
            ]
                \addplot[fill=blue!50, error bars/.cd, y dir=both,
                    y explicit] coordinates {
                    (Greedy(L),0.0190) +- (0,0)
                    (Greedy(C),0.0010) +- (0,0)
                    (Random,0.0258) +- (0,0.0093)
                    (Greedy(Q),0.21) +- (0,0)
                    (Pixie, 0.00843) +- (0,0)
                };

            \end{axis}
        \end{tikzpicture}
        \caption{Avg. Workflow Cost}
        \label{fig:cost_comparison}
    \end{subfigure}
    \caption{QARouter workflow performance comparison of model selection algorithms}
    \label{fig:performance_all_metrics}
\end{figure*}
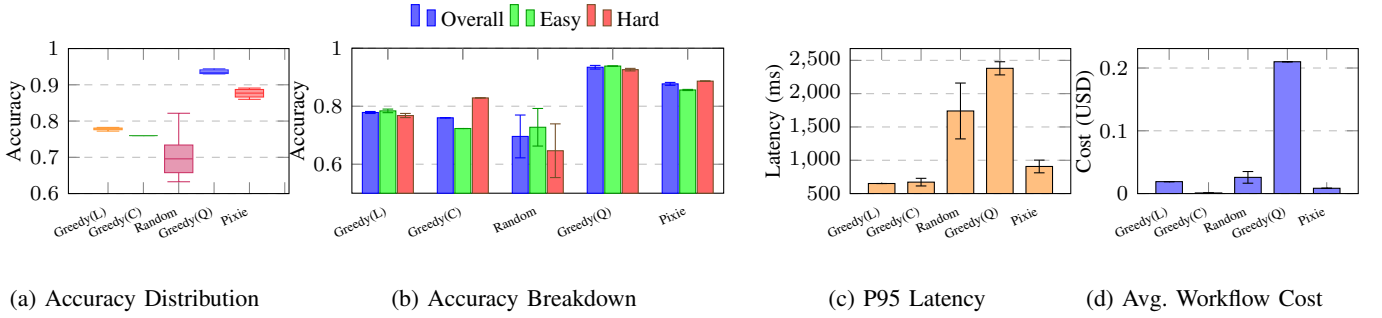

\subsubsection{Infrastructure and Deployment}
Our experimental infrastructure simulates heterogeneous environments representative of the 3D Compute Continuum. For edge deployment, we use an NVIDIA Jetson Orin Nano (8GB RAM) running Triton Inference Server for object detection. Local compute includes a VM with 32-core CPU, 128GB RAM, and NVIDIA RTX 4090 (24GB VRAM) hosting Ollama. Cloud inference uses OpenAI and Anthropic API endpoints. All tasks execute in Docker containers with NVIDIA runtime for GPU isolation. Local models are pre-loaded in GPU memory, enabling model switches in $<$10ms without cold starts.








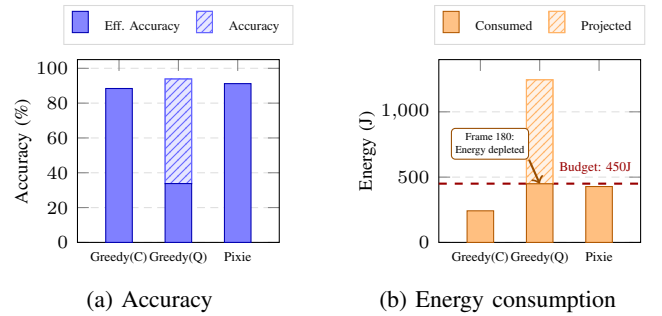
\begin{figure}[H]
\centering

\begin{subfigure}[b]{0.48\columnwidth}
\centering
\begin{tikzpicture}
\begin{axis}[
    ybar stacked,
    width=\linewidth,
    height=4cm,
    bar width=10pt,
    ylabel={Accuracy (\%)},
    ylabel style={font=\scriptsize, yshift=-5pt},
    ymin=0, ymax=105,
    ytick={0,20,40,60,80,100},
    yticklabel style={font=\scriptsize},
    symbolic x coords={GC, GQ, Pixie},
    xticklabels={Greedy(C), Greedy(Q), Pixie},
    xtick=data,
    xticklabel style={font=\tiny},
    enlarge x limits=0.35,
    ymajorgrids=true,
    grid style={densely dashed, gray!25},
    legend style={
        at={(0.5,1.08)},
        anchor=south,
        legend columns=2,
        font=\tiny,
        draw=gray!30,
        fill=white,
        fill opacity=0.95,
        column sep=4pt,
    },
    tick align=inside,
]

\addplot[fill=blue!50, draw=blue!70!black] coordinates {
    (GC, 88.4)
    (GQ, 33.8)
    (Pixie, 91.2)
};

\addplot[
    fill=blue!10,
    draw=blue!60,
    postaction={pattern=north east lines, pattern color=blue!60}
] coordinates {
    (GC, 0)
    (GQ, 60.1)
    (Pixie, 0)
};

\legend{Eff. Accuracy, Accuracy}

\end{axis}
\end{tikzpicture}
\caption{Accuracy}
\label{fig:accuracy_fixed}
\end{subfigure}
\hfill
\begin{subfigure}[b]{0.48\columnwidth}
\centering
\begin{tikzpicture}
\begin{axis}[
    ybar stacked,
    width=\linewidth,
    height=4cm,
    bar width=10pt,
    ylabel={Energy (J)},
    ylabel style={font=\scriptsize, yshift=-5pt},
    ymin=0, ymax=1400,
    yticklabel style={font=\scriptsize},
    symbolic x coords={GC, GQ, Pixie},
    xticklabels={Greedy(C), Greedy(Q), Pixie},
    xtick=data,
    xticklabel style={font=\tiny},
    enlarge x limits=0.35,
    ymajorgrids=true,
    grid style={densely dashed, gray!25},
    tick align=inside,
    legend style={
        at={(0.5,1.08)},
        anchor=south,
        legend columns=2,
        font=\tiny,
        draw=gray!30,
        fill=white,
        fill opacity=0.95,
        column sep=4pt,
    },
]

\addplot[fill=orange!50, draw=orange!70!black] coordinates {
    (GC, 242)
    (GQ, 449)
    (Pixie, 428)
};

\addplot[
    fill=orange!10,
    draw=orange!60,
    postaction={pattern=north east lines, pattern color=orange!60}
] coordinates {
    (GC, 0)
    (GQ, 797)
    (Pixie, 0)
};

\legend{Consumed, Projected}

\draw[dashed, red!60!black, line width=0.8pt]
    ({axis cs:GC,450} -| {rel axis cs:0,0}) -- ({axis cs:GC,450} -| {rel axis cs:1,0});
\node[font=\tiny, red!60!black, anchor=south east]
    at ({axis cs:GC,450} -| {rel axis cs:1,0}) {Budget: 450J};

\coordinate (budgetpoint) at (axis cs:GQ, 450);

\end{axis}

\node[
    draw=orange!60!black,
    font=\fontsize{4}{5}\selectfont,
    text=black!60!black,
    fill=white,
    rounded corners=1.5pt,
    inner sep=1.5pt,
    align=center,
    anchor=east,
] (depleted) at ([xshift=-4pt, yshift=15pt]budgetpoint) {%
    Frame 180:\\
    Energy depleted%
};
\draw[->, thick, orange!60!black] (depleted.south east) -- (budgetpoint);

\end{tikzpicture}
\caption{Energy consumption}
\label{fig:energy_fixed}
\end{subfigure}

\caption{Wildfire workflow performance comparison of model selection algorithms.}
\label{fig:survival}
\end{figure}





\subsection{Wildfire Detection Evaluation}

We evaluate wildfire detection pipeline deployed on edge hardware under energy constraints. This evaluation demonstrates Pixie's ability to achieve full workload processing under fixed energy budgets, while maximizing  detection quality.

\subsubsection{Setup}

\textbf{Candidate Models.} The Detector CAIM has three candidate models from the YOLOv8 family, fine-tuned for fire and smoke detection. Their profiles span 88.6--92.8\% accuracy and 485--2{,}492 mJ per inference, making model selection consequential under a fixed energy budget.


\textbf{Metrics.} We evaluate: (a) \emph{effective accuracy} - correct predictions over the entire workload, penalizing strategies that exhaust energy budgets; (b) \emph{frames processed} - frames completed before energy exhaustion; and (c) \emph{energy consumption}.

\textbf{Dataset.} We use the Wildfire Smoke Dataset~\cite{FireSmoke} containing 4,306 test images with annotated fire and smoke instances across diverse conditions including day/night scenes and varying smoke densities.

\subsubsection{Results}

\cref{fig:survival} presents the results. The results show that static strategies force a trade-off between detection quality and energy consumption. Greedy(Q) achieves 93.9\% per-frame accuracy but exhausts its 450J budget after 180 frames, with a projected energy demand of 1,246J for the full workload. Its effective accuracy drops to 33.8\%. Greedy(C) processes all 500 frames using only 242J but achieves the 88.4\% effective accuracy. Pixie manages this trade-off, processing all 500 frames at 91.3\% effective accuracy, 2.8\% higher than Greedy(C). It successfully combines YOLOv8s (394 frames) and YOLOv8x (106 frames), consuming 438J.




\subsection{QARouter Evaluation}

We evaluate the QARouter workflow under joint SLO constraints on accuracy, latency, and cost. This evaluation demonstrates Pixie's ability to balance multiple SLO dimensions simultaneously.

\subsubsection{Setup}

\textbf{Candidate Models.} The Difficulty Classifier is fixed to DistilBERT (77\% accuracy, ${\sim}$25ms, free). The Simple QA pool comprises three local models (Qwen~2.5, LLaMA~3.2, Gemma2) and one cloud model (GPT-3.5-turbo), spanning 76.9--84.9\% accuracy, 113--717\,ms p95 latency, and \$0--0.042 per 1K inputs. The Complex QA pool comprises four cloud models (GPT-4o-mini, Claude 4 Sonnet, Claude 3 Haiku, Claude 4 Opus), spanning 86.8--96.8\% accuracy, 1229--2180\,ms p95, and \$0.013--1.65 per 1K inputs. The QA solvers exhibit the largest performance and cost variation, so runtime selection focuses on these two CAIMs.


\textbf{Metrics.} We capture workflow execution metrics across three dimensions:
\begin{enumerate*}[label=(\alph*)]
    \item \emph{Quality}: Overall accuracy, accuracy stratified by difficulty (easy/hard questions).
    \item \emph{Latency}: p95 latencies in milliseconds.
    \item \emph{Cost}: Estimated operational cost per request based on provider pricing models.
\end{enumerate*}

\textbf{Dataset.} Experiments use the ARC dataset~\cite{arc} with 3,600 test samples. All experiments were conducted with 5 independent runs using different random seeds to account for variability in routing decisions and model performance.

\textbf{SLO Configuration.} Per-CAIM task SLOs impose quality floors of 75\% accuracy for Simple QA and 85\% for Complex QA, since harder tasks require more reliable models. System SLOs constrain Pixie's runtime selection to p95 latency $\leq$ 1000ms and cost $\leq$ \$0.01 per 600 requests, with budgets decomposed across CAIMs in proportion to profiled per-inference costs. We evaluate end-to-end workflow quality using an 80\% overall success threshold.

\subsubsection{Results}


\begin{figure}[t]
\centering

\begin{subfigure}[t]{0.49\linewidth}
\centering
\begin{tikzpicture}
\begin{axis}[
    width=\textwidth,
    height=3.5cm,
    title style={font=\normalsize\bfseries},
    xlabel={Question Number},
    ylabel={Latency (ms)},
    xlabel style={font=\small},
    ylabel style={font=\small},
    grid=major,
    grid style={dashed, gray!25},
    xmin=0, xmax=600,
    ymin=1400, 
    tick label style={font=\footnotesize},
    xtick={0,200,400,600},
    ytick={2000,3000,4000,5000},
    legend pos=north east,
    legend style={font=\tiny, draw=none, fill=white, fill opacity=0.8},
]
\addplot[
    color=blue!70!black,
    line width=0.6pt,
    mark=none,
    smooth,
] coordinates {
(2, 1997.41) (3, 1836.26) (5, 1708.73) (6, 1741.33) (8, 1685.18)
(9, 1684.60) (10, 1708.42) (13, 1663.46) (14, 1741.94) (15, 1795.31)
(16, 1748.40) (17, 1770.86) (18, 1763.43) (19, 1726.24) (20, 1995.02)
(21, 1809.73) (22, 1902.85) (23, 1883.55) (24, 1901.37) (25, 1855.18)
(26, 1853.58) (27, 1922.81) (28, 1908.49) (29, 1895.74) (30, 1899.45)
(31, 1926.09) (32, 4710.01) (33, 1916.73) (34, 1941.24) (35, 1938.70)
(36, 1928.95) (37, 1962.55) (38, 1973.02) (39, 1985.76) (40, 1999.89)
(41, 2021.03) (42, 2044.16) (43, 2436.11) (44, 2090.34) (45, 2115.21)
(46, 2132.45) (47, 2155.78) (48, 2180.67) (49, 2198.34) (50, 2220.89)
(51, 2245.56) (52, 1548.73) (53, 1769.08) (54, 1519.87) (55, 1506.34)
(56, 1498.92) (57, 1738.74) (59, 1478.56) (60, 1465.89)
(67, 1729.26) (73, 1893.29) (80, 2105.90) (90, 2048.58) (98, 1949.31)
(103, 2150.38) (111, 1518.72) (115, 1843.47) (121, 1638.14) (131, 1795.19)
(137, 1658.83) (143, 2427.96) (148, 1640.40) (159, 1784.31) (164, 1734.07)
(170, 1613.78) (174, 1708.70) (182, 2003.68) (186, 1640.41) (194, 1871.80)
(199, 1846.31) (211, 1634.03) (219, 1645.99) (224, 1871.38) (230, 1693.71)
(237, 1944.45) (248, 1544.76) (253, 1647.90) (260, 1530.66) (268, 1616.32)
(276, 1850.63) (285, 1721.82) (291, 1891.68) (301, 1833.69) (312, 1786.36)
(324, 1684.76) (332, 1742.00) (340, 1744.07) (348, 1737.23) (359, 2060.67)
(366, 1598.61) (373, 2762.74) (383, 1722.28) (393, 1784.40) (403, 1644.45)
(411, 1990.76) (419, 1668.22) (431, 1834.36) (437, 2528.43) (450, 1791.64)
(458, 1742.00) (466, 1875.02) (471, 1942.85) (478, 1718.93) (485, 1986.72)
(496, 2067.74) (505, 1965.44) (510, 2252.86) (518, 1772.27) (525, 2170.58)
(530, 1852.55) (542, 1763.82) (548, 2659.18) (555, 1648.88) (561, 2151.59)
(569, 1817.70) (573, 1740.48) (579, 2100.83) (588, 2046.41) (595, 1945.02)
};
\draw[red!70!black, dashed, line width=0.8pt] (51, 1400) -- (51, 5500);
\node[red!70!black, anchor=south west, font=\scriptsize, fill=white, fill opacity=0.9] 
    at (51, 4000) {Switch Q51};
\end{axis}
\end{tikzpicture}
\caption{Complex QA latencies}
\label{fig:complex_solver}
\end{subfigure}
\hfill
\begin{subfigure}[t]{0.49\linewidth}
\centering
\begin{tikzpicture}
\begin{axis}[
    width=\textwidth,
    height=3.5cm,
    title style={font=\normalsize\bfseries},
    xlabel={Question Number},
    xlabel style={font=\small},
    ylabel style={font=\small},
    grid=major,
    grid style={dashed,gray!25},
    xmin=-10, xmax=600,
    ymin=100, ymax=2400,
    tick label style={font=\footnotesize},
    xtick={0,200,400,600},
    ytick={500,1000,1500,2000},
    legend pos=north east,
    legend style={font=\tiny, draw=none, fill=white, fill opacity=0.8},
]
\addplot[
    color=green!60!black,
    line width=0.6pt,
    mark=none,
    smooth,
] coordinates {
(1, 2297.29) (4, 238.68) (7, 256.48) (11, 256.34) (12, 137.19)
(24, 253.31) (34, 182.32) (40, 440.92) (46, 134.04) (58, 252.12)
(61, 145.23) (62, 156.66) (63, 148.56) (64, 155.78) (65, 160.23)
(66, 158.45) (67, 163.67) (68, 169.89) (69, 172.34) (70, 178.56)
(71, 182.78) (72, 188.90) (73, 195.23) (74, 252.41) (75, 208.67)
(76, 215.89) (77, 223.12) (78, 230.34) (79, 237.56) (80, 244.78)
(81, 252.01) (82, 259.23) (83, 274.40) (84, 273.67) (85, 280.89)
(86, 288.12) (87, 134.13) (88, 302.56) (89, 309.78) (90, 317.01)
(91, 324.23) (92, 331.45) (93, 338.67) (94, 345.89) (95, 353.12)
(96, 133.12) (97, 367.56) (98, 374.78) (99, 382.01) (100, 389.23)
(108, 264.23) (122, 351.88) (129, 245.63) (138, 300.58) (152, 133.29)
(156, 156.84) (169, 228.45) (180, 294.29) (193, 269.02) (203, 133.00)
(207, 180.71) (216, 157.15) (229, 252.49) (239, 258.83) (244, 134.22)
(255, 237.31) (264, 278.32) (271, 276.18) (279, 156.89) (287, 252.83)
(297, 226.61) (303, 273.64) (308, 183.24) (315, 161.51) (321, 157.63)
(326, 214.46) (336, 263.99) (343, 164.11) (352, 161.88) (356, 155.98)
(365, 133.69) (374, 288.65) (381, 292.73) (387, 276.25) (395, 255.42)
(401, 263.12) (408, 255.22) (415, 156.01) (422, 132.87) (429, 157.82)
(438, 269.23) (443, 156.95) (449, 133.64) (455, 159.39) (464, 157.48)
(475, 159.75) (486, 253.64) (492, 156.02) (498, 157.65) (507, 232.14)
(517, 274.02) (529, 260.76) (537, 188.10) (541, 157.69) (551, 161.69)
(564, 254.19) (574, 266.04) (582, 161.63) (590, 158.54) (601, 256.17)
};
\draw[red!70!black, dashed, line width=0.8pt] (58, 100) -- (58, 2400);
\node[red!70!black, anchor=south west, font=\scriptsize, fill=white, fill opacity=0.9] 
    at (58, 1800) {Switch Q58};
\end{axis}
\end{tikzpicture}
\caption{Simple QA latencies}
\label{fig:simple_solver}
\end{subfigure}

\vspace{0.6em}

\begin{subfigure}[b]{0.99\columnwidth}
\centering
\begin{tikzpicture}
\begin{axis}[
    width=\linewidth,
    height=3.8cm,
    grid=major,
    grid style={dashed, gray!30},
    xlabel={Question Number},
    ylabel={Cumulative Cost (\$)},
    title style={font=\normalsize\bfseries},
    xlabel style={font=\small},
    ylabel style={font=\small},
    xmin=0, xmax=600,
    ymin=0, ymax=0.05,
    xtick={0,100,200,300,400,500,600,700,800,869},
    yticklabel style={/pgf/number format/fixed, /pgf/number format/precision=3},
    legend pos=north west,
    legend style={font=\tiny, fill=white, fill opacity=0.8, draw=gray},
]
\addplot[
    color=blue!70!black,
    line width=1.0pt,
    smooth,
    mark=none,
] coordinates {
    (1,0.000006)
    (20,0.005199)
    (40,0.026694)
    (51,0.033307)
    (58,0.033391)
    (80,0.033651)
    (100,0.033891)
    (120,0.034131)
    (140,0.034371)
    (160,0.034611)
    (180,0.034851)
    (200,0.035091)
    (220,0.035331)
    (240,0.035571)
    (260,0.035811)
    (280,0.036051)
    (300,0.036291)
    (320,0.036531)
    (340,0.036771)
    (360,0.037011)
    (380,0.037251)
    (400,0.037491)
    (420,0.037731)
    (440,0.037971)
    (460,0.038211)
    (480,0.038451)
    (500,0.038691)
    (520,0.038931)
    (540,0.039171)
    (560,0.039411)
    (580,0.039651)
    (600,0.039891)
    (620,0.040131)
    (640,0.040371)
    (660,0.040611)
    (680,0.040851)
    (700,0.041091)
    (720,0.041331)
    (740,0.041571)
    (760,0.041811)
    (780,0.042051)
    (800,0.042291)
    (820,0.042531)
    (840,0.042771)
    (860,0.043011)
    (869,0.043232)
};

\addplot[color=red, mark=*, mark size=2.5pt, only marks] coordinates {(51,0.033307)};
\addplot[color=violet, mark=*, mark size=2.5pt, only marks] coordinates {(58,0.033391)};

\draw[red!60, dashed, line width=0.8pt] (axis cs:51,0) -- (axis cs:51,0.033307);
\draw[violet, dashed, line width=0.8pt] (axis cs:58,0) -- (axis cs:58,0.033391);

\node[red!70!black,anchor=south west, font=\scriptsize] at (axis cs:51,0.034) 
    {\parbox{3.5cm}{\raggedright Switch 1 (Q51):\\Claude Opus $\rightarrow$ GPT-4o-mini}};
\node[color=violet,anchor=south west, font=\scriptsize] at (axis cs:58,0.0170) 
    {\parbox{3.5cm}{\raggedright Switch 2 (Q58):\\Gemma2:2b $\rightarrow$ Llama3.2:3b}};
\node[anchor=north east, font=\scriptsize, text width=4.5cm] at (axis cs:600,0.016) 
    {\textit{Note: Early steep cost growth (Q1--50) stabilizes after switches.}};
\end{axis}
\end{tikzpicture}
\caption{Cumulative cost across all questions with switch markers.}
\label{fig:cumulative_cost}
\end{subfigure}
\caption{Latency and cost behavior with model switches}
\label{fig:solver_latencies_and_cost}
\end{figure}
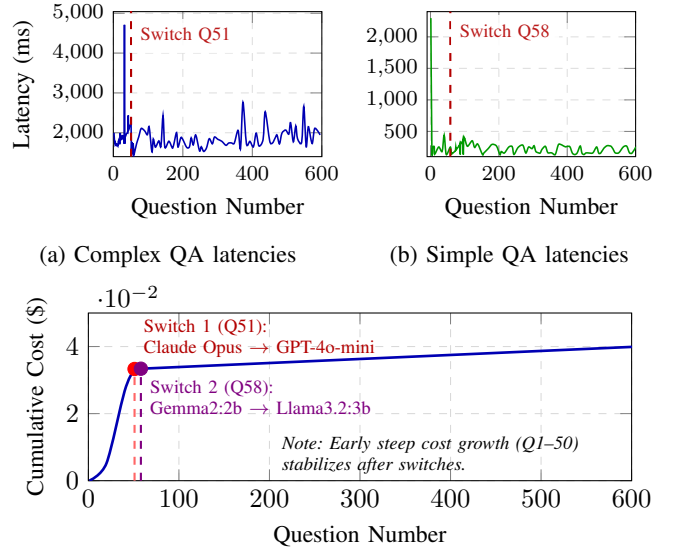

\textbf{Comparison to Static Baselines.} Figure~\ref{fig:performance_all_metrics} shows Pixie achieves 87.71\% accuracy, approaching Greedy-Quality's 93.44\% while using 99\% less cost (\$0.008 vs \$0.81) and being 4.5× faster. Compared to Greedy-Cost's minimal \$0.001 spend, Pixie accepts 8× higher cost to gain 11.7\% in accuracy, crossing the SLO threshold that Greedy-Cost misses at 76\%. Against Greedy-Latency, Pixie improves accuracy by 10\% and reduces cost by 56\%, while satisfying latency SLO.

Pixie's runtime model selection is the only strategy satisfying all three SLOs simultaneously. Static baselines optimize single metrics at the expense of others: Greedy(Q) exceeds budget by 21×, while Greedy(C) and Greedy(L) fall 4--6\% below accuracy requirements.

\textbf{Model Switching Impact.} \cref{fig:solver_latencies_and_cost} illustrates Pixie's dynamic adaptation capabilities through runtime model switching under relaxed SLOs (p95 $<$ 2500ms, cost $<$ \$0.05). These SLOs were chosen to force Pixie to initially select high-quality models, and then perform cost-driven switches at Q51 and Q58 when approaching budget constraints. This demonstrates Pixie's ability to maintain SLO compliance through runtime adaptation.

\section{Related Work}\label{sec:related}

We review prior work across two areas relevant for enabling runtime model selection: (1) compound AI programming abstractions and (2) model selection algorithms.

\subsection{Compound AI Programming Abstractions}


Existing body of work has proposed various declarative programming abstractions. DSPy \cite{DSPY} abstracts language model pipelines as parameterized modules with natural language signatures, eliminating manual prompt engineering through automatic optimization. LangChain \cite{chase2022langchain} provides a framework for building LLM applications through composable abstractions. However, both DSPy and LangChain require developers to specify computational pipeline structure and model assignments. Murakkab~\cite{MURAKKAB} provides a declarative abstraction where workflow tasks are expressed as natural language descriptions and mapped to AI models via an LLM-based orchestrator. However, this mapping still requires the orchestrator to bind specific AI models to tasks before execution.

Although these approaches successfully address important aspects of compound AI complexity, they couple workflow definition to specific AI models, preventing adaptation. Compoundable AI Models address this gap by decoupling workflow specification from underlying AI models, thus, enabling runtime adaptation through model selection.

\subsection{Model Selection Algorithms}

Recent advances in model selection for compound AI systems have achieved significant performance improvements. Approaches such as LLMSelector~\cite{chen2025optimizingmodelselectioncompound} optimize model allocation by iteratively selecting modules and allocating models with the highest module-wise performance estimated by LLM diagnosers. Beyond LLM-specific optimization, model selection algorithms address heterogeneous model types across vision, audio, and classification tasks \cite{OPTSURVEY}. EdgeAdaptor \cite{EdgeAdaptor} tackles joint model selection and resource provisioning for edge DNN inference serving through MILP approach, while INFaaS~\cite{INFaaS} generates a set of model variants for a specified AI model and selects between them. 


However, existing approaches focus on single model inference serving or perform model selection during deployment time that remain fixed throughout workflow execution. Our work addresses these limitations through CAIM abstraction coupled with Pixie runtime model selection. Contrary to current approaches, Pixie leverages runtime monitoring data to make informed selection decisions during execution, maintaining SLO compliance.


\section{Conclusion}\label{sec:conclusion}

PLAIground demonstrates that SLO compliance in Compound AI systems can be maintained through runtime model selection. The CAIM abstraction decouples workflow logic from model implementations via Task, Data, and System Contracts, making runtime model switching possible without modifying application code. Building on this, Pixie continuously selects the most suitable candidate model per CAIM by monitoring observed resource metrics against SLO thresholds, upgrading or downgrading assignments as conditions change. Together, these allow Pixie to be the only evaluated strategy satisfying accuracy, latency, and cost SLOs simultaneously, achieving up to 91.3\% accuracy while reducing costs by up to 21x over static baselines. Future work includes developing advanced model selection algorithms leveraging predictive decision making and multi-objective optimization and novel SLO decomposition algorithms focusing on accuracy estimation for Compound AI workflows.


\section*{Acknowledgment}

This research was partially funded by the EU's Horizon Europe Research and Innovation Program as part of the NexaSphere project (GA No. 101192912).

\bibliographystyle{IEEEtran} 
\bibliography{references} 

\end{document}